\documentclass[preprint,11pt]{elsarticle}

\usepackage[top=2.5cm,bottom=2.5cm,left=2.6cm,right=2.6cm,a4paper]{geometry}
\usepackage{graphicx}
\usepackage{dcolumn}
\usepackage{bm}
\usepackage{epstopdf}
\usepackage{mathrsfs}
\usepackage{amssymb,amsfonts,latexsym}
\usepackage{amsmath,bbold}
\usepackage{color}
\newcommand{\beq}{\begin{eqnarray}}
\newcommand{\eeq}{\end{eqnarray}}
\newcommand{\nn}{\nonumber\\}

\newcommand{\rmd}{{\rm d}}
\newcommand{\rme}{{\rm e}}

\def\btheta{{\bm \theta}}
\def\q{{\bm q}}
\def\p{{\bm p}}

\def\k{{\boldsymbol k}}

\def\r{{\boldsymbol r}}

\def\u{{\boldsymbol u}}

\newcommand{\del}{\partial}

\newcommand{\abar}{\bar{\alpha}}

\begin{document}
\begin{frontmatter}

\title{\Large Angular distribution of medium-induced QCD cascades }

\author[ipht]{Jean-Paul Blaizot}
\ead{jean-paul.blaizot@cea.fr}
\author[ipht]{Leonard Fister}
\ead{leonard.fister@cea.fr}
\author[ipht]{Yacine Mehtar-Tani}
\ead{yacine.mehtar-tani@cea.fr}

\address[ipht]{%
Institut de Physique Th\'eorique, CEA Saclay, F-91191 Gif-sur-Yvette,
France

}%

\begin{abstract}
We provide a complete description of the angular distribution of gluons in a medium-induced QCD cascade. We identify two components in the distribution, a soft component dominated by soft multiple scatterings, and a hard component   dominated by a few hard scatterings. The typical angle that marks the boundary between these two components is determined analytically as a function of the energy of the observed gluon and the size of the medium. 
We construct the complete solution (beyond the diffusion approximation) in the regime where multiple branchings dominate the dynamics of the cascade in the form of a power series in the number of collisions with the medium particles. The coefficients of this expansions are related to the moments of the distribution in the diffusion approximation and are determined analytically. The angular distribution may be useful in phenomenological studies of jet shapes in heavy-ion collisions.  
\end{abstract}

\begin{keyword}
Perturbative QCD \sep Heavy-Ion collisions \sep Jet-quenching
\end{keyword}

\end{frontmatter}
\begin{flushright}
\footnotesize{PACS numbers: 12.38.-t,24.85.+p,25.75.-q}
\end{flushright}

\section{Introduction}
\label{sec:intro}

Jet measurements at the LHC represent a highlight of the heavy-ion program. High energy jets have the potential to probe the nature of the hot and dense matter produced in heavy ion collisions, the so-called quark-gluon plasma. The data reveal indeed that jets are strongly attenuated as they traverse a quark-gluon plasma, and a careful study of this attenuation can help pinning down various properties of the plasma, such as some of  its transport  coefficients.  Particularly interesting for the present work is the detailed analysis of the missing energy observed in inbalanced dijet events \cite{Aad:2010bu,Chatrchyan:2011sx,CMS2014,Chatrchyan:2012ni}.  The reconstruction of the energy and angular distribution of jets particles around the dijet axis has been achieved, and evidence has been obtained that the missing energy in the away side jet is found in the form of soft particles radiated at very large angles from the direction of the jet.  

On the theory side, there is a large dispersion in the predictions of the various jet-quenching models, which calls for a more complete,  first principle, theory of jets in heavy-ion collisions. This paper reports on progress in this direction. The angular distribution that we determine may be a useful ingredient to be implemented in  Monte Carlo event generators that are developed by several groups \cite{Schenke:2009gb,Zapp:2011ya,Renk:2008pp,Armesto:2009fj}. 
The present work builds on the  probabilistic equation that describes  the in-medium QCD cascade, and which was derived recently \cite{Blaizot:2013vha}. It controls the evolution with the size of the medium, $L$, of the inclusive  distribution of partons produced in the cascade as a function of the energy and the transverse momentum of the observed gluon. By solving this equation, we can determine the angular distribution of the energy that is radiated by the leading particle. This is the aim of the present work.  

The first moment of the angular distribution (or equivalently of the transverse momentum distribution) was discussed recently \cite{Blaizot:2014ula,Kurkela:2014tla}. Three regimes can be identified as a function of the energy $\omega$ of the observed gluon: when $\omega\lesssim E$, with $E$ the energy of the leading particle,  we are measuring the original parton, whose transverse momentum is not altered by radiation and is determined solely by multiple scattering with the medium. The typical transverse momentum that it acquires is $\langle k_\perp^2\rangle \sim\hat q L$, where $\hat q$ is the so-called quenching parameter and corresponds to a diffusion coefficient in transverse momentum space. As we shall shorty see, $\hat q $ determines not only the physics of transverse momentum broadening but also that of medium-induced parton branchings.  The second regime is that for which the energy $\omega$ of the observed gluon is much smaller than $E$ but larger than the characteristic scale  $\omega_s=\alpha^2_s\hat q L^2$, where $\alpha_s$ is the strong coupling constant, which marks the onset of branchings. This gluon is most likely radiated by the leading parton, and can be produced anywhere inside the medium. Its typical transverse momentum is $\langle k_\perp^2\rangle \sim \hat q L/2$, where $L/2$ is the average 
time spent by the radiated gluon in the medium. Finally, in the multiple branching regime, $\langle k_\perp^2\rangle \sim \hat q t_\ast(\omega)$, with $t_\ast$ the typical time spent in the medium by the measured parton: 
\beq
t_\ast(\omega)\sim \frac{1}{\alpha_s}\sqrt{\frac{\omega}{\hat q }}\,.
\eeq 
In this regime, the typical transverse momentum is independent of the size of the medium. 

\begin{figure}[!ht]
\begin{center}
\includegraphics[width=14cm]{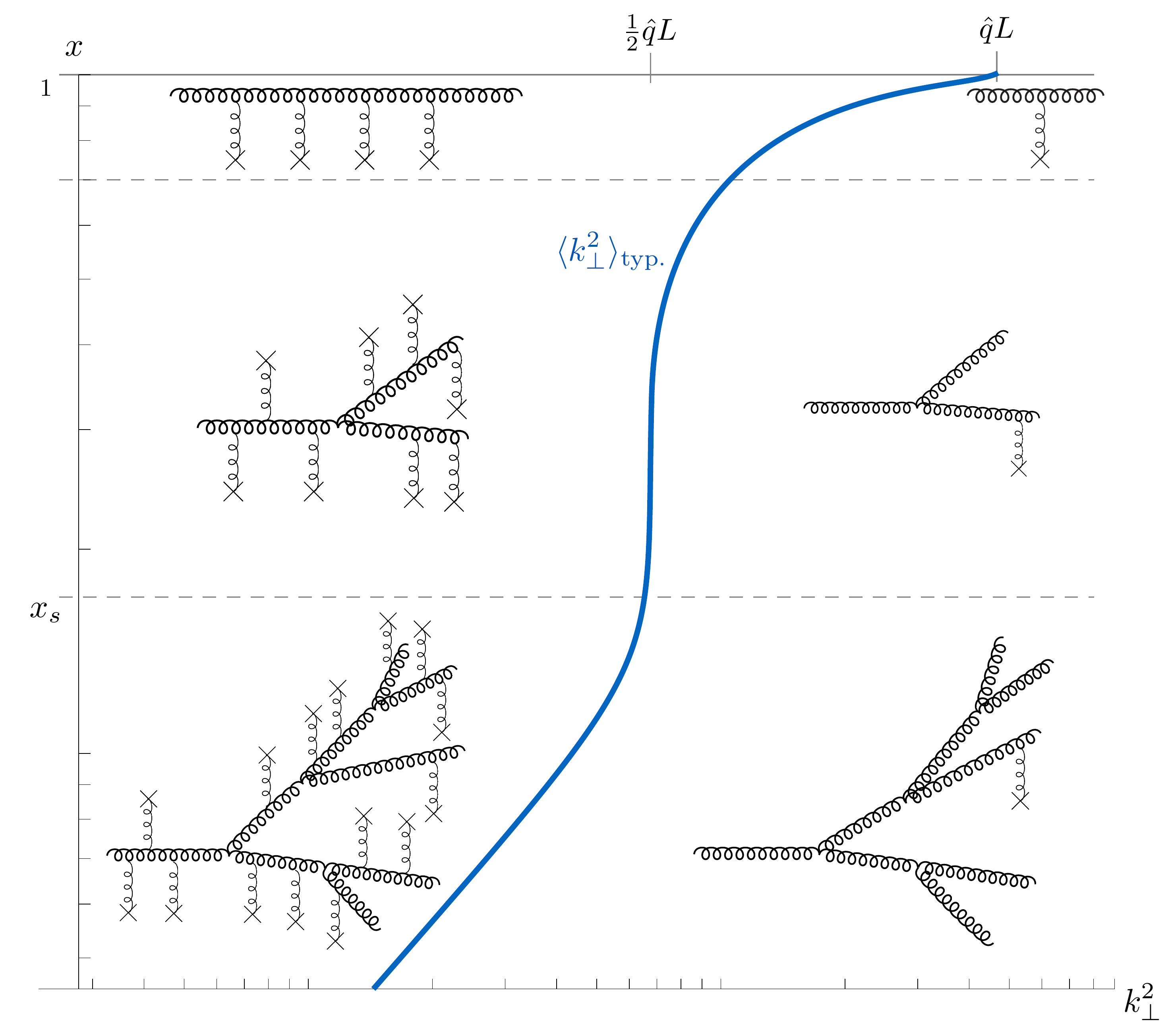}
\caption{(Color online.) The various regimes of the physical processes (branching and scattering) that accompany the propagation of a fast parton in a dense medium. The thick (blue) line separates the regime of single hard, large angle, scattering, from that of soft, small angle, multiple scattering. Accordingly, the soft component of the angular distribution lies on the  left of the blue line, the hard component on the right. The line $\omega/E\equiv x=x_s$ separates the region dominated by single branching from that of multiple branchings. The horizontal line below $x=1$ indicates the region where the leading particle propagates without splitting, and merely suffers  momentum broadening due to its collisions with the constituents of the medium. \label{fig1}}
\end{center}
\end{figure}

In addition to the three regimes that we have just discussed, the character of the angular (or transverse momentum) distribution changes depending on whether $k_\perp$ is smaller or larger than the typical transverse momentum $ \langle k_\perp^2\rangle $ that we have identified  for each of the energy regimes: for $k_\perp \ll \langle k_\perp^2\rangle $ multiple elastic scatterings dominate the dynamics, in the opposite case, i.e.,  $k_\perp \gg \langle k_\perp^2\rangle $, only a single scattering determines the transverse momentum distribution and the distribution drops as $k_\perp^{-4}$.  We shall refer to these two parts of the angular distribution as to the soft and hard components, respectively. The pattern of the various regimes if illustrated in  Fig.~\ref{fig1}.

In this paper, we complete the analysis undertaken with the study of the first moment, and provide a complete description of the  angular distribution.
 In section \ref{sec:cascade}, we recall the basic equation that governs the evolution with the size of the medium of the inclusive distribution of gluons, as a function of the energy $\omega$ and the angle $\theta$ of the observed gluon.  We discuss two limiting cases. In the  first case,  the transverse momentum is integrated out, leaving us with the equation for the energy distribution. The other case corresponds to freezing the branching processes, and therefore on momentum broadening, with the identification of two regimes, that of multiple soft scatterings and that of hard single scattering.  In section \ref{sec:solution}, we construct the angular distribution in the three regimes of energies that are discussed above.  The analytic calculation of the moments of the soft component of the distribution are presented in Appendix A.

\section{The equation for the in-medium QCD cascade}
\label{sec:cascade}

We consider the cascade of radiated gluons that is generated by a high energy gluon\footnote{The extension to the case of a high energy quark  is straightforward}, with energy $E$, that  propagates through a quark-gluon plasma. We focus on the medium-induced radiation, as governed by the BDMPSZ mechanism \cite{Baier:1996kr,Zakharov:1996fv,Baier:2001yt},  and ignore here effects due to vacuum radiation. In this approach, both gluon splitting and momentum broadening are controlled by a single parameter $\hat q$, called the jet-quenching parameter. Thus, for instance, the average transverse momentum acquired by a parton through its collisions with the plasma constituents  is $\langle k_\perp^2\rangle\sim \hat q L$, with $L$ the length of the medium crossed by the parton. The BDMPSZ mechanism takes into account the Landau-Pomeranchuck-Migdal  (LPM) effect and provides the dominant contribution to the gluon spectrum for gluon frequencies $\omega_{\rm BH}\lesssim \omega\lesssim \omega_c$, where $\omega_c\sim \hat q L^2$ is the maximum energy that can be taken 
away by a single 
gluon (the present analysis assumes\footnote{We do not anticipate major qualitative changes when  $\omega_c<E$ \cite{exactsol,Fister:2014zxa}.} that $\omega_c\gtrsim E$). 
The lower limit is that of (Bethe-Heitler) incoherent emissions, and is reached when the branching time is of the order of the mean free path between successive collisions. The branching time for a gluon with energy $\omega$ is given by $\tau_{\rm br}(\omega)\sim\sqrt{\omega/\hat q}$. It is associated with a transverse momentum scale $k_{\rm br}(\omega)\sim(\omega\hat q)^{1/4}$, and an emission angle $\theta_{\rm br}(\omega)\sim(\hat q/\omega^3)^{1/4}$.  

We  are interested in the inclusive  gluon distribution
\beq
D(x,\k,t)\equiv (2\pi)^2 x \frac{dN}{dxd^2\k}\,,
\eeq
where  $x=\omega/E$ is the energy fraction,  and $\k$ the  transverse momentum, of the gluon observed at some time $t$ along the cascade, with the maximum value of $t$ equal to the length $L$ of the medium.  As was shown in \cite{Blaizot:2013vha}, $D(x,\k,t)$ obeys the following  integro-differential equation 
\beq\label{evol-eqk}
\frac{\del}{\del t}D(x,\k,t)=\frac{1}{t_\ast}\int \rmd z\, {\cal K}(z)\left[\frac{1}{z^2}\sqrt{\frac{z}{x}} D\left(\frac{x}{z},\frac{\k}{z},t\right)-\frac{z}{\sqrt{x}} D\left(x,\k,t \right)\right]+\int_\q{\cal C}(\q) D(x,\k-\q,t)\,, \nn
\eeq
where  we introduced in the last term  a 
shorthand  notation for the transverse momentum integrations: $\int_\q\equiv \int \rmd^2\q/(2\pi)^2$. This will be used  throughout the paper. The kernel ${\cal K}(z)$  in the gain and loss terms (respectively the first and second term in the r.h.s. of Eq.~(\ref{evol-eqk})), can be written as \cite{Blaizot:2012fh}
\beq\label{Kdef}
{\cal K}(z)=\,\frac{[f(z)]^{5/2}}
{[z(1-z)]^{3/2}}\,,
\qquad f(z)\equiv 1-z+z^2\,.
\eeq
It collects contributions from  the $z$ dependence of the actual branching time (left out in the definition\footnote{The actual branching time for offsprings carrying fractions $z$ and $1-z$ of the initial energy is $\sqrt{z(1-z)}t_\ast$.} of $t_\ast$, Eq.~(\ref{stop-time}) below), and from the leading order splitting function $P_{gg}(z)=N_c [f(z)]^2/z(1-z)$ where $N_c$ is the number of colors (we restrict our discussion to purely gluonic cascades). Equation~(\ref{evol-eqk}) combines two distinct physical effects: i) the branching of gluons, which preserves the angles but change the number of particles, and is described by the term proportional to $1/t_\ast$; ii)  the momentum broadening due to collisions with the plasma constituents, which preserves the number of particles, and is described  by the last term.

Two important approximations are involved in the derivation of Eq.~(\ref{evol-eqk}). First, the typical duration of the branching process is assumed to be small compared to the total time spent by the gluon in the medium. This allows to treat  the branchings as effectively instantaneous \cite{Blaizot:2012fh}. Second, the transverse momentum broadening that takes place during a branching is ignored (corrections involving the small transverse momentum induced during the splitting can be absorbed in corrections to $\hat q$~\cite{Blaizot:2013vha,Blaizot:2014bha}  (see also \cite{Liou:2013qya,Iancu:2014kga} for related discussions of radiative corrections to $p_\perp$ broadening). The branching is then treated as effectively collinear: After the splitting, the two new gluons carry fractions $z$ and $1-z$ of both the initial energy and the initial transverse momentum.   Thus when a gluon of energy $\omega_0$, and transverse momentum $\k_0$ splits into a gluon of energy $\omega_1$, and transverse momentum $\k_1$ and another gluon of 
energy $\omega_2$, and transverse momentum $\k_2$, we have $\omega_1=z\omega_0$, $\k_1=z\k_0$ (and similarly for the other gluon, with $z\to 1-z$). It follows that $\theta_0=k_0/\omega_0=k_1/\omega_1=\theta_1$, the angle is preserved in the splitting. On the other hand, if $z\ll 1$, $k_1\ll k_0$, there is a degradation of the transverse momentum at each splitting along the cascade. The competition between this degradation of the transverse momentum that accompanies each splitting, with the accumulation of transverse momentum form collisions with the medium constituents, plays an important role in determining the form of the momentum distribution. \\

The quantity 
\beq\label{stop-time}
\frac{1}{t_\ast}\equiv\frac{\bar\alpha}{\tau_{_{\rm br}}(E)} = \bar\alpha\sqrt{\frac{\hat q}{E}}\,,
\eeq
with $ \bar\alpha\equiv {\alpha_s N_c}/{\pi}$, 
is the basic rate of the branching processes. It depends on the energy $E$ of the incoming parton,  and on the medium through the jet-quenching parameter $\hat q$.  We shall soon verify that  $t_\ast$ is the time at which most of the energy of the incoming parton  has been radiated into soft gluons, and for this reason it is sometimes referred to as the stopping time. There are two basic control parameters in the problem: the energy $E$ and the size $L$ of the medium.  We expect several regimes, depending on the values of these control parameters. The characteristic time $t_\ast$ is related to the energy $E$ as follows
\beq
E=\bar\alpha^2 \hat q t_\ast^2\,.
\eeq
The frequency that characterizes the onset of the  multiple branching regime is given by an analogous formula
\beq
\omega_s=\bar\alpha^2 \hat q L^2=\abar^2\omega_c\,,
\eeq
where $\omega_c=\hat q L^2$. 
Gluons that are observed with an energy $\omega\lesssim \omega_s$ are likely to come from multiple branchings. Indeed the probability for emitting a gluon with energy $\omega$ in a distance $L$ is $\bar\alpha L/\tau_{\rm br}(\omega)=\bar\alpha L\sqrt{\hat q/\omega}$. This is of order unity when $\omega\sim \omega_s$ and larger when $\omega<\omega_s$. Now, by comparing the two equations above, we see that if $L>t_\ast$, then $\omega_s> E$ and  one is always in the regime of multiple branching, since all radiated gluons have frequencies  $\omega<E<\omega_s$. If on the other hand, $L<t_\ast$, the situation that we shall consider in this paper, then several regimes appear depending on whether we consider gluons with $\omega<\omega_s$ or $\omega>\omega_s$. These various regimes are illustrated in Fig.~\ref{fig1}.\\

The last term in Eq.~(\ref{evol-eqk}) describes momentum broadening, with the collision kernel given by 
\beq\label{collisionkernel}
{\cal C}(\q)=w(\q)-(2\pi)^2\delta(\q)\int_{\q'} w(\q')\,,\qquad w(\q)=\frac{16 \pi^2 \alpha^2_s N_c n}{\q^4}\,,
\eeq
where $n$ is the density of scatterers (which we assume to be independent of $t$ for simplicity). The quantity $w(\q)\rmd^2\q \rmd t$ can be interpreted as the probability that the particle acquire a transverse momentum $\q$ during $\rmd t$: $w(\q)$ is proportional to the elastic scattering cross section $\rmd^2\sigma_\text{el}/\rmd^2\q$ of the fast parton with the constituents of the medium, while ${\cal C}(\q)$ is proportional to the so-called dipole cross section $\sigma(\q)$, ${\cal C}(\q)=-N_cn \sigma(\q)/2$. 
The jet-quenching parameter is related to the scattering rate. To logarithmic accuracy,  it is given by
\beq\label{qhat0}
\hat q(\p^2) = \int_\q \q^2 w(\q)\simeq 4\pi \alpha_s^2 N_c \,n  \, \ln\frac{\p^2}{m_D^2}\,,
\eeq
where  the Debye screening mass $m_D$  regulates the singular behavior of the collision kernel at small momenta, while the upper cutoff is the typical momentum of the observed gluon. This formula makes sense, as does the definition of $\hat q$, only in the regime dominated by soft multiple scatterings. As we shall verify later this implies 
\beq\label{boundary}
\p^2 \lesssim \hat q(\p^2) L\,.
\eeq

Insight into the general behavior of the solution of Eq.~(\ref{evol-eqk}) can be gained by considering limiting cases where only branchings, or only scatterings take place. We start by the former case.

\subsection{The energy distribution }
\label{sec:energy-dist}

By integrating Eq.~(\ref{evol-eqk})  over the transverse momentum, one obtains an evolution equation for the energy density $D(x,t)$:
\beq\label{evol-eq-0}
\frac{\del}{\del t}D(x,t)=\frac{1}{t_\ast}\int \rmd z\, {\cal K}(z)\left[\sqrt{\frac{z}{x}} D\left(\frac{x}{z},t\right)-\frac{z}{\sqrt{x}} D\left(x,t\right)\right]\,.
\eeq
The scattering term vanishes upon integration over $\k$. 
Note that the function $D(x,t)$ has support only for $0\le x\le 1$,
which limits the first $z$-integral in Eq.~(\ref{evol-eq-0}) to $x< z < 1$. Note also that the potential endpoint singularities at $z=1$ in the gain and loss terms cancel. 

This equation can be solved exactly in the case where, in the kernel ${\cal K}(z)$ in Eq.~(\ref{Kdef}), $f(z)$ is set equal to unity \cite{Blaizot:2013hx}.  
 This simplification preserves the  singular behavior of the kernel near $z=0$ and $z=1$,  which determines the qualitative features of the solution.  We shall use this exact solution from now on. For the initial condition $D(x,t=0)=\delta(1-x)$, this solution reads
\beq\label{Dexact}
  D(x,t)\,=\,\frac{\tau}{\sqrt{x}(1-x)^{3/2}}\exp \left(-\pi\frac{\tau^2}{1-x}\right)\,,\qquad \tau\equiv \frac{t}{t_\ast}\,.
  \eeq  
 The essential singularity at $x=1$  can be understood as a Sudakov suppression factor  \cite{Baier:2001yt}
 (i.e. the vanishing of the probability to emit no gluon in any finite time).  
 One can easily verify  on this explicit solution the interpretation of  $t_\ast$ as a stopping time: after a time $t\simeq t_\ast$ most ($\gtrsim 98\%$) of the energy is to be found in the form of radiated soft ($x\lesssim 0.1$) gluons. Aside from this exponential factor, the solution has another remarkable property: for $x\ll 1$,  $D(x,t)$ factorizes into a function of time and a function of $x$
 \beq\label{scalingspectrum}
  D(x,t)\sim \frac{1}{\sqrt{x}}\, \tau {\rm e}^{-\pi \tau^2},\qquad (x\ll 1)\,.
\eeq
The fact that the spectrum keeps the same  $x$-dependence 
when $t$ keeps increasing reflects the fact that the energy flows to  $x=0$ without
accumulating  at any finite value of $x$, a property reminiscent of wave turbulence \cite{Blaizot:2013hx}. The complete, energy conserving, solution involves a contribution $\propto\delta(x)$ whose coefficient grows with time as $1-{\rm e}^{-\pi\tau^2}$.  Note that the onset of  the regime dominated by multiple branchings is not directly visible on the scaling spectrum (\ref{scalingspectrum}): there is no change of behavior when  $x\lesssim x_s=\omega_s/E$. 

The short time behavior of the solution (\ref{Dexact}) will be useful in the foregoing discussion. This  can be obtained iteratively by inserting in the right hand side of Eq.~(\ref{Dexact}) the leading particle solution $D^{(0{\rm b})}(x,t)=\delta(1-x)$, where the upper script $(0{\rm b})$ refers to the solution with zero branching. Alternatively, one can just simply expand Eq.~(\ref{evol-eq-0}) for small $\tau$, and obtain, for $x$ not too close to 1, 
the solution corresponding to one branching, 
\beq\label{D1b}
D^{(1{\rm b})}(x,t)\,=\,\frac{\tau}{\sqrt{x}(1-x)^{3/2}}=\frac{t}{t_\ast} x{\cal K}(x)\,.
\eeq
The last equality indicates the relation of this approximate distribution with the BDMPSZ spectrum.

\subsection{The transverse momentum distribution for a single particle}
\label{sec:pT-dist}

We turn now to the scattering term. We may formally isolate its effects by letting $t_\ast\to\infty$, thereby effectively switching off the contributions of the gluon branching.   The resulting distribution can then be written as 
\beq\label{LP-1}
D(x,\p,t)=\delta(1-x) {\cal P}(\p,t)\,,
\eeq
with ${\cal P}(\p,t)$ the probability for the leading gluon to acquire a transverse momentum $\p$ during its propagation through the plasma for a duration $t$. This quantity obeys the equation
\beq\label{equaforP}
\frac{\del}{\del t}{\cal P}(\p,t)=\int_\q{\cal C}(\q){\cal P}(\p-\q,t)\,, 
\eeq
 with the collision kernel given by Eq.~(\ref{collisionkernel}). 
The  equation (\ref{equaforP}) can be solved, formally, by Fourier transform. Setting ${\cal C}(\q)=\int \rmd^2\r\, {\rm e}^{i\q\cdot\r}{\cal C}(\r)$, we get
\beq\label{Pcoord}
{\cal P}(\p,L)=\int \rmd^2\r \exp\left[-i\p\cdot\r+\int_{0}^{L} \rmd t\,{\cal C}(\r)\right]\,.
\eeq
Although this integral cannot be calculated analytically, its main features can be easily obtained. For instance, one may expand the second exponential in powers of the dipole cross section and give the result an interpretation in terms of multiple scattering \cite{Blaizot:2004wu}. On may then identify two components in the distribution ${\cal P}(\p,L)$: a soft component dominated by multiple scattering, in which case the multiple scattering expansion cannot be truncated to a finite number of terms, and a hard tail populated by rare events with single hard scatterings, for which the leading term in the multiple scattering series is sufficient to obtain  an accurate determination of ${\cal P}$. 

In order to quantify the separation between these two regimes, it is useful to consider an approximate expression of the Fourier transform ${\cal C}(\r)$:
\beq\label{sigmadipole}
{\cal C}(\r)=16\pi^2 \alpha^2_s N_c n\int_\q\frac{\rme^{i\q\cdot\r}-1}{\q^4}\simeq -\pi \alpha^2_s N_c n\, \r^2\ln\left( \frac{1}{\r^2 m_D^2}  \right)=- \frac{1}{4} \hat q(\r^{-2})\, \r^2\,,
\eeq
where we have used the expression (\ref{qhat0}) for $\hat q$, and chosen $1/\r^2$ as the upper cut-off scale. Multiple scattering start to dominate when $L{\cal C}(\r)$ in Eq.~(\ref{Pcoord}) becomes of order unity. We recover the condition (\ref{boundary}) since  in Eq. (\ref{Pcoord}) $\r^{-1}\sim \p$.

In the regime dominated by multiple soft scatterings,  the momentum transfer $\q$ in a collision with medium particles is small compared to $\hat q L$,  i.e., $m_D^2\ll \hat q L$. Equivalently, in this regime,  typically $\q\ll \p$ in Eq.~(\ref{equaforP}), and one can transform this equation into a Fokker-Planck equation: 
\beq\label{P-mom-FP}
\frac{\del}{\del t}{\cal P}(\p,t) = \frac{1}{4}\,\left(\frac{\del}{\del \p}\right)^2\,\Big[ \hat q(\p^2) \,{\cal P}(\p,t)\Big]\,,
\eeq
with the jet quenching parameter $\hat q(\p^2)$ playing the role of a (momentum dependent) diffusion coefficient. It is given, to logarithmic accuracy,  by Eq.~(\ref{qhat0}).
Hence, the typical transverse momentum squared acquired by a particle after a a time $t$ of propagation in the medium is $Q^2_s(t) = \hat q t$. 
In the approximation where one ignores the (logarithmic) dependence of $\hat q$ on the dipole size $\r$, an approximation often  referred to as the ``harmonic approximation'', one can easily solve the diffusion equation (\ref{P-mom-FP}). Assuming $n$, and hence $\hat q$, to be independent of $t$ for simplicity, one gets
\beq\label{Gaussdistr}
{\cal P}(\p,L)=\frac{4\pi}{Q^2_s(L)}\, \exp\left[-\frac{\p^2}{Q^2_s(L)}\right]\,.
\eeq

As already emphasized, the diffusion picture is valid in the regime dominated by multiple scattering, and  holds for $\p^2\lesssim \hat q L$. Larger transverse momenta can be achieved through a single hard scattering. The expression for ${\cal P}$ corresponding to a single hard scattering is  easily obtained from Eq.~(\ref{equaforP}), and reads
 \beq\label{Phard}
{\cal P}(\p,L)\approx (2\pi)^2 \delta(\p) \left[ 1-L\int_\q w(\q) \right]+\frac{16\pi^2 \alpha^2_s N_c n L}{\q^4}\,,
\eeq
Note that because of this high momentum tail, the momentum distribution does not admit moments beyond the leading one (i.e., the integral of the distribution, related to conservation of probability, and the integral weighed by $|\p|$).

\subsection{The evolution equation for the angular distribution }
\label{sec:evol-eq-angle}

As we have mentioned, the splittings are treated as collinear, meaning that there is no deflection of particles caused by the splittings.   It is then more convenient to follow the evolution of angles rather than that of the transverse momenta along the cascade. Accordingly we transform the transverse momentum distribution into an angular distribution, setting 
\beq
D(x,\btheta)\equiv (2\pi)^2 x\frac{\rmd N}{\rmd x\rmd^2\btheta}\,,
\eeq
where 
\beq
\btheta\equiv\frac{\k}{\omega}=\frac{\k}{xE}\,.
\eeq
Note that $\btheta$ is a 2-dimensional vector collinear to $\k$, whose (small) magnitude equals the polar angle of the emitted gluon with respect to the initial direction of the  leading particle. This distribution is normalized as follows
\beq\label{nomalization}
\int \frac{\rmd^2 \btheta}{(2\pi)^2} \,D(x,\btheta)=D(x)\,.
\eeq
In this new variable, Eq. (\ref{evol-eqk}) reads 
\begin{eqnarray}
\frac{\del}{\del t}D(x,\btheta,t)&=&\frac{1}{t_\ast}\int \rmd z\, {\cal K}(z)\left[\sqrt{\frac{z}{x}} D\left(\frac{x}{z},\btheta,t \right)-\frac{z}{\sqrt{x}} D\left(x,\btheta ,t\right)\right]\nn
&&+\int \frac{\rmd^2\btheta'}{(2\pi)^2}{\cal C}(\btheta',x) D(x,\btheta-\btheta',t)\,,
\label{evol-eq-ang0}
\end{eqnarray}
with (see Eq.~(\ref{collisionkernel})) 
\beq\label{calCx}
{\cal C}(x,\btheta)=(xE)^2{\cal C}(\q)= \frac{16\pi^2 \alpha^2_s N_c n }{(xE)^2} \left[\frac{1}{\btheta^4}- \delta(\btheta)\int \frac{\rmd^2\btheta' }{\btheta'^4} \right]\,.
\eeq
The locality (in angle) of the splitting term reflects the collinearity of the splitting. 
In the diffusion approximation, i.e., $\btheta' \ll \btheta $ in Eq.~(\ref{evol-eq-ang0}), the equation  (\ref{evol-eq-ang0}) reduces to 
\begin{eqnarray}\label{evol-eq-diff0}
\frac{\del}{\del t}D(x,\btheta,t)&=&\frac{1}{t_\ast}\int \rmd z\, {\cal K}(z)\left[\sqrt{\frac{z}{x}} D\left(\frac{x}{z},\btheta,t \right)-\frac{z}{\sqrt{x}} D\left(x,\btheta,t \right)\right] \nn
&&+\frac{1}{4(xE)^2} \,  \left(\frac{\partial}{\partial \btheta}\right)^2 \, \left[\hat q \, D(x,\btheta,t)\right]\,,
\end{eqnarray}
with $\hat q\simeq  4\pi\alpha_s^2 N_c n\ln ({\btheta^2}/{\btheta_D^2})$ and 
 $\btheta_D\equiv m_D/\omega=m_D/(xE)$. \\

It useful to proceed in Fourier space. We set
\beq\label{FT}
D(x,\u,t)=\int \frac{\rmd^2\btheta}{(2\pi)^2} D(x,\btheta,t)\,\rme^{-i\btheta\cdot\u}\,,
\eeq
and  get 
\beq\label{evol-eq-ang}
\frac{\del}{\del t}D(x,\u,t)=\frac{1}{t_\ast}\int \rmd z\, {\cal K}(z)\left[\sqrt{\frac{z}{x}} D\left(\frac{x}{z},\u,t\right)-\frac{z}{\sqrt{x}} D\left(x,\u,t\right)\right]+{\cal C}(x,\u) D(x,\u,t)\,, 
\eeq
where 
\beq
{\cal C}(x,\u)=\int \frac{\rmd^2\btheta}{(2\pi)^2}{\cal C}(x,\btheta)\,\rme^{-i\btheta\cdot\u}\,.
\eeq
This equation  is the same as that, Eq.~(\ref{evol-eq-0}), satisfied by $D(x,t)$, except for the last term which plays here the role of source term. We can then write the solution   as follows
\beq\label{evol-eq-ang2a}
D(x,\u,L)=\int_0^L \rmd t\int_x^1\frac{\rmd y}{y} \,D\left(\frac{x}{y},\frac{L-t}{\sqrt{y}}\right) {\cal C}(y,\u)D(y,\u,t)\,.
\eeq
The function $D(x/y,(L-t)/\sqrt{y})$, given by the solution (\ref{Dexact}) of Eq.~(\ref{evol-eq-0}), plays here the role of a Green's function (and will be often referred to  as such in the foregoing discussion). In Eq.~(\ref{evol-eq-ang2a}) it takes the explicit form 
 \beq\label{Greenfunction}
 D\left(\frac{x}{y}, \frac{L-t}{\sqrt{y}}\right)=\frac{L-t}{t_\ast \sqrt{x} \left( 1-\frac{x}{y} \right)^{3/2} } 
 \exp\left[- \frac{\pi}{y-x} \left( \frac{L-t}{t_\ast} \right)^2 \right]\,.
 \eeq
 We can easily verify  the following convolution property
 \beq\label{convolution}
 \int_x^1 \frac{\rmd y}{y}\, D\left(\frac{x}{y},\frac{L-t}{\sqrt{y}}\right)\,D(y,t)= D(x,L)\,.
 \eeq
 In the foregoing discussion we shall need this Green's function in various regimes. For $L-t\ll t_\ast$, it is simply the delta function $y\delta(y-x)$, corresponding to the propagation without any branching. For one branching, $ D\left(\frac{x}{y}, \frac{L-t}{\sqrt{y}}\right)$ can be approximated by the solution (\ref{D1b}). Finally, in the multiple scattering regime, the following integral will be useful
 \beq
 \label{integralG}
  \int_0^\infty\rmd t' D\left(\frac{x}{y}, \frac{t'}{\sqrt{y}}\right)=\frac{\sqrt{y}\, t_\ast }{  2\pi \sqrt{x/y} \sqrt{1-x/y} }\,.
 \eeq
 
\section{Determining the angular distribution of gluons}
\label{sec:solution}

Solving Eq.~(\ref{evol-eq-ang0}) exactly is difficult. In this section, we shall obtain an analytic representation of the solution in the various regimes indicated in Fig.~\ref{fig1}, where appropriate approximations can be made:   $x\simeq 1$ which corresponds to the leading particle; $x_s\ll x\ll 1$ which corresponds to the primary gluon radiation, and finally the regime  $x\ll x_s$ dominated by  multiple branching. 
Part of the difficulty in solving Eq.~(\ref{evol-eq-ang0}) comes from the fact that, as we have already emphasized, the angular distribution has two distinct components: a hard component, corresponding to large angles produced by single hard scatterings, and a soft component that can be obtained as the solution of the diffusion equation (\ref{evol-eq-diff0}). 
Besides, these two components are strongly modified by gluon branching. 
 
The soft component admits moments, which is not the case for the hard component. The characteristic angle that marks the boundary between the soft and the hard components depends on $x$, i.e., on the amount of branching. It can be estimated by calculating the mean squared angle of the soft component, and we proceed to its determination in the next subsection.

\subsection{The mean squared angle $\langle \theta^2\rangle$ }
\label{sec:angsquare}

We define the typical angle squared 
\beq
\langle \theta ^2 \rangle = \frac{M_1(x,L)}{D(x,L)}\,,
\eeq
where $M_1(x,L)$ is the first moment of the angular distribution obtained in the diffusion approximation, Eq.~(\ref{evol-eq-diff0}), 
\beq
M_1(x,L)=\int \frac{\rmd^2\btheta}{(2\pi)^2}\, \btheta^2 D(x,\btheta,L)\,.
\eeq
This moment obeys the following equation \cite{Blaizot:2014ula}
\beq\label{evol-eq-1}
\frac{\del}{\del t}M_1(x,t)=\frac{1}{t_\ast}\int \rmd z\, {\cal K}(z)\left[\sqrt{\frac{z}{x}} \,M_1\left(\frac{x}{z},t\right)-\frac{z}{\sqrt{x}} M_1\left(x,t\right)\right]+\frac{N^2\hat q}{(xE)^2} \,D(x,t)\,.
\eeq
It can be solved formally by multiplying it  by the Green's function (\ref{Greenfunction}) and integrating over  $t$ and $y$. We then obtain the following integral representation for $\langle \theta^2 \rangle$
\beq\label{ang-sol-1a}
\langle \theta^2 \rangle =\frac{ \hat q}{E^2}\int_0^L \rmd t\int_x^1 \frac{\rmd y}{y^3}\, D\left(\frac{x}{y},\frac{L-t}{\sqrt{y}}\right)\, \frac{D(y,t)}{D(x,L)}\,.
\eeq
We now proceed to the approximations that are valid in the different regimes of interest. \\

When $x\simeq 1$, we can write
\beq
\int_x^1 \frac{\rmd y}{y^3}\, D\left(\frac{x}{y},\frac{L-t}{\sqrt{y}}\right)\, D(y,t)&\simeq & \int_x^1 \frac{\rmd y}{y}\, D\left(\frac{x}{y},\frac{L-t}{\sqrt{y}}\right)\,D(y,t)= D(x,L)\,,
\eeq
where we have used, in the first equality, that $y^3\simeq y $ in the integral measure,  and the property (\ref{convolution}) in the last equality.  
We then get the simple result, which corresponds to the momentum broadening of the leading particle,
\beq\label{ang-sol-2}
\langle \theta ^2 \rangle =\frac{ \hat qL}{E^2}= \theta_s^2(1,L),\qquad \theta_s^2(x,L)\equiv \frac{ \hat qL}{\omega^2}=\frac{ \hat qL}{x^2E^2}\,.
\eeq

Let us turn now to the regime of  primary gluon radiation, i.e., $x_s\ll x\ll 1$. It is easily verified that, when $x\ll 1$, the dominant contribution to $M_1(x,t)$ is obtained 
when the first $D$ in Eq.~(\ref{ang-sol-1a}) is $D^{(0{\rm b})}$ and the second one $D^{(1{\rm b})}$. That is, the dominant contribution corresponds to that of the momentum broadening of the radiated gluon from the time of its emission. The other contribution, corresponding to  the momentum broadening of the leading particle before the splitting, is suppressed by a factor $x^2$. We get then, keeping the two contributions, 
\beq\label{ang-sol-3}
\langle \theta ^2 \rangle =\left( 1+\frac{1}{x^2}  \right)\frac{ \hat qL}{2E^2}=\frac{1}{2}\left( 1+x^2  \right)\theta_s^2(x,L)\,.
\eeq

Let us now discuss the third regime, $x\ll x_s$, which is characterized by the multiple branchings. 
In the small $x$ region, namely for $x\ll x_s$ multiple branchings require a non perturbative treatment. 
To do that we return to the expression (\ref{Greenfunction}) of the Green's function and  note that 
the integral over $y$ in Eq.~(\ref{ang-sol-1a}) is weighed towards values $x \lesssim y\ll 1$. The Green's function (\ref{Greenfunction}) decays exponentially as a function of its time argument, except in the region 
\beq
L-t \lesssim \sqrt{y}t_\ast\ll  \sqrt{x_s}t_\ast\ = L\,.
\eeq
On the other hand, $D(y,t)$  decays over a time scale of order $t_\ast\gg L$. On that time scale the Green's function appears as a sharply peaked function of its time argument. 
In order to extract the leading behavior in this regime, we can thus set $t\sim L$ in $D(y,t)$  in Eq.~(\ref{ang-sol-1a}), and integrate freely the Green's function over $t'=L-t$ from 0 to $\infty$. By using Eq.~(\ref{integralG}), one gets then
\beq\label{D-it1}
M_1(x,\btheta,L)&=&\frac{\hat q}{E^2}\int_x^1\,\frac{\rmd y}{y} \frac{\sqrt{y}\, t_\ast }{  2\pi \sqrt{x/y} \sqrt{1-x/y} }\frac{1}{y^2} D(y,L)\nn &\approx& \frac{\hat q}{E^2}D(x,L)\int_x^1\,\frac{\rmd y}{2\pi y^3}\frac{\sqrt{y}}{\sqrt{1-x/y}}\nn
&\approx& \frac{\hat q}{(xE)^2}D(x,L)\frac{t_\ast \sqrt{x}}{4}\,,
\eeq
where in the second line, we have used $D(y,t)\approx \sqrt{x/y}\,D(x,t)$ valid in the scaling regime $x\lesssim y\ll 1$, and in the last one we have set the lower bound of the integration to zero and used
\beq
\int_0^1\rmd u\frac{u}{\sqrt{u(1-u)}}=\frac{\pi}{2}\,.
\eeq
Finally, we get
\beq
\langle \theta ^2 \rangle = \frac{1}{4\bar \alpha}\left [\frac{\hat q}{(xE)^3}\right]^{1/2}\equiv \theta^2_\ast(x)\,.
\eeq

In summary, the boundary of the soft part of the angular distribution is given at large $x$ by $\theta_s^2(x,L)$ which  depends on the size $L$ of the medium, and at small $x$ by $\theta_\ast^2(x)$ which is independent of  $L$. The line that separates the two components of the distribution in Fig.~\ref{fig1} reflects qualitatively  this behavior. 

\subsection{The leading parton: $x\simeq 1$} 
Recall that we focus in this analysis on leading gluons  that have sufficient energy to escape the medium without being completely absorbed. That is, we assume $t_\ast \gg L$. 
The distribution (\ref{Dexact}) exhibits then a peak near $x\sim 1$ that corresponds to the leading particle, and a radiation spectrum growing as $x^{-1/2}$ for small decreasing values of $x$. 
When $x\sim 1$, the distribution takes then the factorized form
\beq\label{dist-x0}
D(x,\btheta,L)\simeq {\cal P} (\btheta,L) \, D(x,L)\,,
\eeq 
where $D(x,L)$ is energy distribution given by Eq.~(\ref{Dexact}) (for $x\lesssim 1$), and  ${\cal P} (\btheta,L)$ solves the equation that derives from  Eq.~(\ref{equaforP}) after making the change of variables $\btheta=\k/(xE)$, namely, 
\beq\label{equaforP-2}
\frac{\del}{\del t}{\cal P}(\btheta,t)=\int \frac{\rmd^2\btheta'}{(2\pi)^2}{\cal C}(x,\btheta'){\cal P}(\btheta-\btheta',t)\,,
\eeq
with the initial condition ${\cal P}(\btheta,0) = \delta(\btheta)$.
The solution of this equation has essentially been given already in the previous section. In the regime of multiple scatterings, i.e., for $\theta^2\lesssim \theta_s^2(L,x)$ it reads (see Eq.~(\ref{Gaussdistr}))
\beq\label{Gaussdistr2}
{\cal P}(\btheta,L)=\frac{4\pi}{\theta_s^2(x,L)}\exp\left[-\frac{\btheta^2}{\theta_s^2(x,L)}\right]\,.
\eeq
In the opposite case, $\theta \gg \theta_s(x,L)$, a single hard scatterings dominates the distribution (cf. Eq.~(\ref{Phard})) and we have (see Eq.~(\ref{Phard}))
\beq\label{1-scat-P}
{\cal P}(\btheta,L)=\frac{16\pi^2 \alpha^2_s N_c nL}{(xE)^2 \, \btheta^4}\sim \frac{\theta^2_s(L,x)}{\btheta^4}\,.
\eeq
\subsection{Single BDMPSZ  radiation: $x_s\ll x\ll1$} 

Let us now discuss the angular distribution of the primary gluon emissions off the leading particle. 
In this regime the energy distribution is given by the leading order BDMPSZ distribution, 
\beq
D(x,L)\simeq \frac{L}{t_\ast \sqrt{x}}\,,
\eeq 
and  we expect also  the angular distribution to be given by a single radiation that undergoes multiple scatterings. To see that, we look for the single branching contribution in Eq.~(\ref{evol-eq-ang0}), 
\begin{eqnarray}
\frac{\del}{\del t}D(x,\btheta,t)&=&\frac{1}{t_\ast}\int \rmd z\, {\cal K}(z)\left[\sqrt{\frac{z}{x}} D^{(0{\rm b})}\left(\frac{x}{z},\btheta,t \right)-\frac{z}{\sqrt{x}} D^{(0{\rm b})}\left(x,\btheta,t \right)\right]\nn
&&+\int \frac{\rmd^2\btheta'}{(2\pi)^2}{\cal C}(\btheta',x) D(x,\btheta-\btheta',t)\,, 
\label{evol-1b}
\end{eqnarray}
where the zero branching contribution is given by 
\beq
D^{(0{\rm b})}\left(x,\btheta,t \right)=\delta(1-x) \, {\cal P}(\btheta,t)\,.
\eeq
By performing the integration over $z$ in Eq.~(\ref{evol-1b}), one gets
\beq
\frac{\del}{\del t}D(x,\btheta,t)=\frac{1}{t_\ast}\,x{\cal K}(x)\, {\cal P}(\btheta,t)+\int \frac{\rmd^2\btheta'}{(2\pi)^2}{\cal C}(\btheta',x) D(x,\btheta-\btheta',t)\,.
\eeq
By using the Green's function ${\cal P} (x,\btheta, t-t_0)$ that obeys the equation
\beq\label{Grenn-P}
\frac{\del}{\del t}{\cal P} (x,\btheta,t-t_0)-\int \frac{\rmd^2\btheta'}{(2\pi)^2}{\cal C}(\btheta',x) {\cal P} (x,\btheta-\btheta', t-t_0)= \delta(\btheta)\delta(t-t_0)\,,
\eeq
one obtains
\beq\label{evol-1b-2}
D(x,\btheta,L)=\int_0^L\frac{\rmd t}{t_\ast}\,   \int \frac{\rmd^2\btheta'}{(2\pi)^2}{\cal P} (x,\btheta-\btheta', L-t)   \,x{\cal K}(x)\,{\cal P}(\btheta',t)\,.
\eeq
Note that  ${\cal P}(1,\btheta,t)\equiv {\cal P}(\btheta,t)$.\\
This equation can understood as follows: the leading parton broadens from $0$ to $t$ with a probability ${\cal P}(\btheta',t)$, emits a soft gluon at  time $t$, which propagates from $t$ to $L$ and whose distribution broadens according to ${\cal P} (x,\btheta-\btheta', L-t)$. 
For sufficiently small $x$, the angular deviation of the the radiated gluon is larger than that of the leading parton. Accordingly one can neglect $\btheta'$ in the argument of the first $\cal P$, which allows us to perform trivially  the integral over $\btheta'$,  
\beq
 \int \frac{\rmd^2\btheta'}{(2\pi)^2}{\cal P}(\btheta',t)=1\,.
\eeq
We finally get, 
\beq\label{evol-1b-3}
D(x,\btheta,L)\simeq \int_0^L\frac{\rmd t}{t_\ast}\,   {\cal P} (x,\btheta,L-t)   \,x{\cal K}(x)\,. 
\eeq
One can check that the moments of the angular distribution that are calculated in  Appendix A using a different technique, coincide indeed with those of the distribution (\ref{evol-1b-3}).

\subsection{Multiple branchings: $x\ll x_s$} 

We are left now with the fully non-perturbative regime, i.e., $x\ll x_s$, where, in the soft region,  both multiple scatterings and multiple branchings must be resummed. 
In order to solve Eq.~(\ref{evol-eq-ang0}) in this regime, we look for a solution as a power series in the number of scatterings. 
Our starting point is Eq.~(\ref{evol-eq-ang2a}), which we   rewrite here for convenience
\beq\label{evol-eq-ang2}
D(x,\u,L)&=&\int_0^L \rmd t\int_x^1\frac{\rmd y}{y} \,D\left(\frac{x}{y},\frac{L-t}{\sqrt{y}}\right) {\cal C}(y,\u)D(y,\u,t)\,.
\eeq
This form of the equation allows us to resum easily the multiple scattering series. To that aim, we set
\beq\label{seriesforD}
D(x,\u,t)=\sum_{n=0}^\infty D_n(x,\u,t)\,,
\eeq
with $D_n$ of order ${\cal C}^n$. 
We obtain then, from Eq.~(\ref{evol-eq-ang2}),
\beq\label{evol-eq-ang3}
D_n(x,\u,L)= \int_0^L \rmd t \int_x^1\frac{\rmd y}{y}D\left(\frac{x}{y},\frac{L-t}{\sqrt{y}}\right){\cal C}(y,\u) D_{n-1}(y,\u,t)\,, 
\eeq
with  $D_0(x,\u,t)=D(x,t)$, and 
\beq\label{sigmadipole2}
{\cal C}(x,\u)=\int \int \frac{\rmd^2\btheta}{(2\pi)^2}{\cal C}(x,\btheta){\rm e}^{-i\btheta\cdot\u}\simeq -\frac{\pi \alpha^2_s N_c n}{(xE)^2}\u^2\ln\left( \frac{1}{\u^2 \theta_D^2}  \right),\qquad \theta_D\equiv \frac{m_D}{xE}\,.
\eeq
 To proceed we note that the dependence of ${\cal C}(y,\u)$ on the variables $y$ and $\u$ occurs in the form
\beq
{\cal C}(y,\u)\sim \frac{\u^2}{y^2}\, \ln\left(\frac{y^2E^2}{\u^2 m_D^2}\right)\,,
\eeq
so that  
\beq
\frac{{\cal C}(y,\u)}{{\cal C}(x,\u)}&=&  \frac{x^2}{y^2}\,\frac{\ln\left(\frac{y^2E^2}{\u^2m_D^2}\right)}{\ln\left(\frac{x^2E^2}{\u^2m_D^2}\right)}= \frac{x^2}{y^2}\left[1+\frac{\ln\left(\frac{y^2}{x^2}\right)}{\ln\left(\frac{x^2E^2}{\u^2m_D^2}\right)}\right]\,.
\eeq
The logarithm in the denominator has typically a large argument: this is because in the multiple scattering regime, the typical angle $\theta \sim 1/u_\perp$ is achieved by multiple collisions deflecting the particle by an angle $\theta_D=m_D/(xE)\ll 1/u_\perp$. On the other hand,  we have already observed that the integral over $y$ in Eq.~(\ref{evol-eq-ang3})  is dominated by values of $y$  larger but of the order of $x$. Therefore, the ratio of logarithms is generically small and can be neglected. Thus, in a first approximation, one can write 
\beq
{\cal C}(y,\u)&\simeq& {\cal C}(x,\u) \frac{x^2}{y^2}\,.
\eeq
This allows us to separate the variables in Eq.~(\ref{evol-eq-ang3}), which becomes then
\beq\label{evol-eq-ang4}
D_n(x,\u,L)\simeq {\cal C}(x,\u)  \int_0^L \rmd t \int_x^1\frac{\rmd y}{y}D\left(\frac{x}{y},\frac{L-t}{\sqrt{y}}\right) \frac{x^2}{y^2} D_{n-1}(y,\u,t)\,.
\eeq
The structure of this integral ressembles that encountered when computing the mean transverse momentum squared, Eq.~(\ref{ang-sol-1a}). We may then proceed similarly as in Eq.~(\ref{D-it1}) in order to extract the leading behavior, and set $t\sim L$ in $D_{n-1}(y,\u,t)$  in Eq.~(\ref{evol-eq-ang4}), and integrate freely the Green's function over $t'=L-t$ from 0 to $\infty$. Then, proceeding by recursion, we postulate the following form of the solution
\beq\label{evol-eq-ang5}
D_n(x,\u,L)&=&  c_n \Big[{\cal C}(x,\u) t_\ast(x) \Big]^n D(x,L)\,,
\eeq
where, for  $x\ll x_s$,  $D(x,L)\sim 1/\sqrt{x} \,f(L)$ and
\beq
t_\ast(x)=\frac{1}{4\bar\alpha}\sqrt{\frac{Ex}{\hat q}}\,,
\eeq
The unknowns in Eq.~(\ref{evol-eq-ang5}) are the coefficients $c_n$. These are obtained by substituting the expression (\ref{evol-eq-ang5}) in Eq.~(\ref{evol-eq-ang3}). We obtain then the recursion  formula
\beq\label{evol-eq-ang6}
c_n= c_{n-1} \, t^{-1}_\ast(x)  \int_0^\infty \rmd t' \int_0^1\frac{\rmd y}{y}D\left(\frac{x}{y},\frac{t'}{\sqrt{y}}\right) \left(\frac{x}{y} \right)^{\frac{3n+1}{2}}\,,
\eeq
or, using the explicit value of the integral (\ref{integralG}) 
\beq
c_n &= & c_{n-1} \, \frac{1}{2\pi}\int_0^1 \rmd u \frac{u^{\frac{3n-2}{2}}}{\sqrt{1-u}}\,.
\eeq
The integration over $u$ yields 
\beq
\int_0^1 \rmd u \frac{u^{\frac{3n-2}{2}}}{\sqrt{1-u}}=B\left(\frac{3n}{2},\frac{1}{2}\right)=\frac{\Gamma\left(\frac{3n}{2}\right)\Gamma\left(\frac{1}{2}\right)} {\Gamma\left(\frac{1+3n}{2}\right)}\,.
\eeq
Recalling that $\Gamma(1/2)=\sqrt{\pi}$ and the fact that $c_0=1$, we finally get
\beq\label{evol-eq-ang7}
c_n &= & \prod_{m=1}^n \frac{2}{ \sqrt{\pi} }\frac{\Gamma\left(\frac{3m}{2}\right)} {\Gamma\left(\frac{1+3m}{2}\right)}\,.
\eeq
These coefficients entirely determine the Fourier transform $D(x,\u,L)$ (see Eqs.~(\ref{seriesforD}) and (\ref{evol-eq-ang5})), from which the angular distribution $D(x,\theta,L)$ can in principle be deduced by performing the inverse Fourier transform. This can only be done numerically and remains a delicate procedure.  

\begin{figure}[!ht]
\begin{center}
\includegraphics[width=10cm]{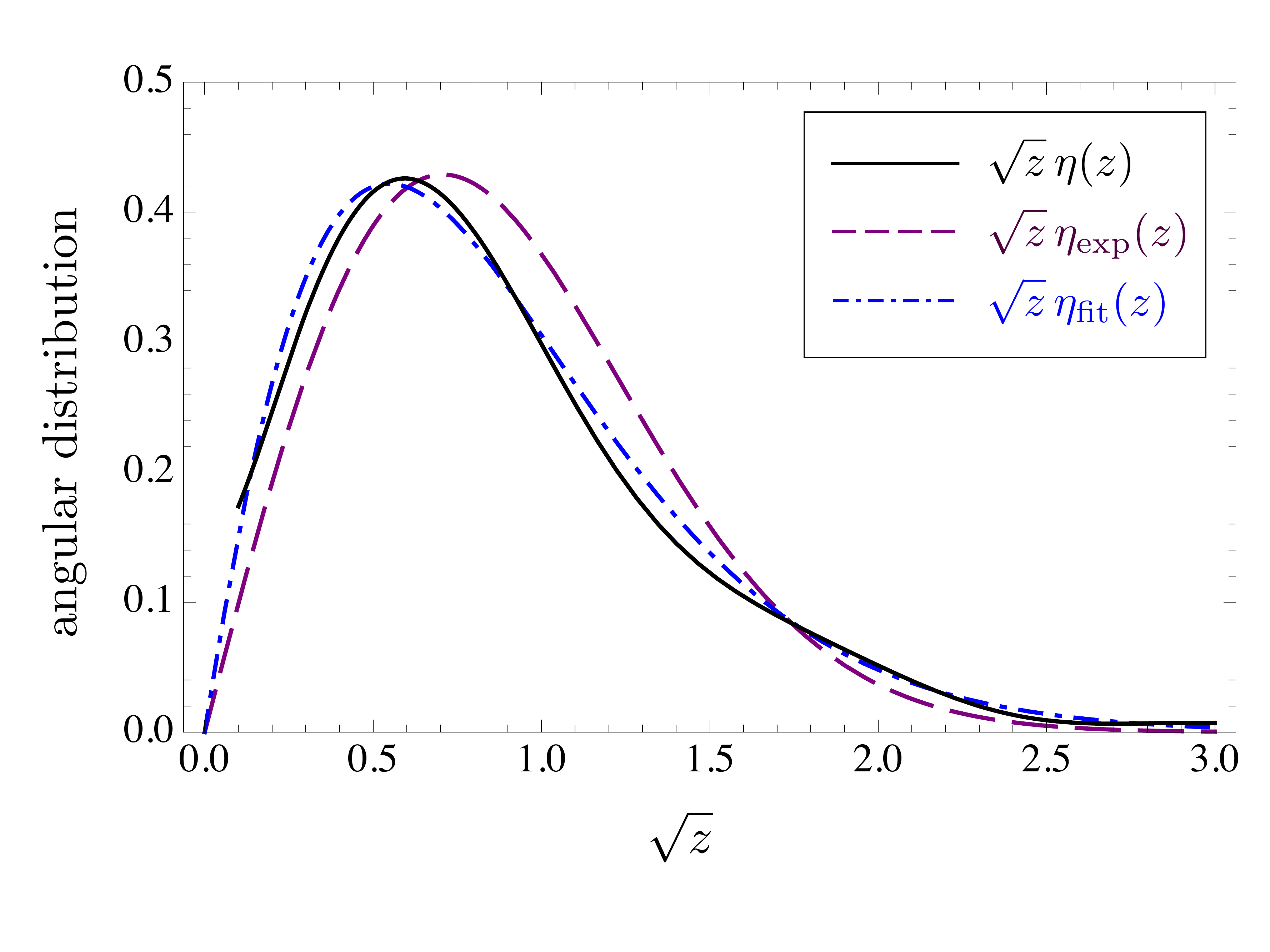}
\caption{(Color online.) The angular distribution  (\ref{scal-eta-2}) of a gluon in the multiple branching regime (solid black line)  compared to an exponential distribution with identical first two moments (dashed purple line). It can be fitted with a single parameter function (cf. Eq.~(\ref{fit})) inspired by the large angle limit (\ref{large-theta}) (dashed-dotted blue line).  }
\label{fig2}
\end{center}
\end{figure}

We have carried it in  the harmonic approximation,
\beq
{\cal C}(x,\u)\approx \frac{1}{4}\frac{\hat q }{(xE)^2}\, \u^2\,,
\eeq
where  the logarithmic dependence of the dipole cross-section on $\u$ and $x$ is ignored. This approximation, which is equivalent to the diffusion approximation, does not allow us to explore the tail of the distribution for very large angles (see below), but it allows us to determine the distortion of the main peak.  
In this approximation, we simply get
\beq\label{sol-diff}
D(x,\u,L)&=& D(x,L) \sum_{n=0}^{\infty} c_n \left[-\frac{1}{4} \theta^2_\ast(x)\, \u^2 \right]^n\,,
\eeq
where we have used the fact that $ \hat q t_\ast(x)= (xE)^2 \,\theta^2_\ast(x)$. 
After performing an inverse Fourier transform we can write the angular distribution as follows,
\beq
D(x,\theta,L)=\frac{4\pi}{\theta^2_\ast(x)}\,  \eta\left(\frac{\theta^2}{\theta^2_\ast(x)}\right)\,D(x,L)\,,
\eeq 
where the scaling function $\eta$ is given by
\beq\label{scal-eta}
 \eta\left(\frac{\theta^2}{\theta^2_\ast(x)}\right)&=& \frac{1}{4\pi} \theta^2_\ast(x) \int \rmd^2\u \, \rme^{i\btheta\cdot \u } \, \sum_{n=0}^{\infty} c_n \left[-\frac{1}{4} \theta^2_\ast(x)\, \u^2 \right]^n\,.
\eeq
After integrating over the azimuthal angle and renaming the variables,  $z=\theta^2/\theta^2_\ast(x)$ and $2\alpha=|\u| \theta_\ast(x)$,  we can rewrite Eq.~(\ref{scal-eta}) as 
\beq\label{scal-eta-2}
\eta(z) &=&  \int_0^\infty \rmd \alpha \, J_0(2\sqrt{z\alpha})  \, \sum_{n=0}^{\infty} c_n \left(-\alpha^2 \right)^n\,,
\eeq
where $J_0$ is a Bessel function. 
Note that the property (\ref{nomalization}) implies that
\beq\label{eq:eta_norm}
\int_0^\infty \rmd z\, \eta(z) =1\,.
\eeq
In Fig.~\ref{fig2} we have plotted the angular distribution $\eta(z)$ in the multiple branching regime.  For the numerical evaluation we have computed the first 500 terms in the series (\ref{sol-diff}). We have checked firstly, that the first three moments of the distribution agree with the moments computed analytically from Eq.~(\ref{a-N-coef}) and, secondly, that the large angle behavior matches the asymptotic limit, see Eq.~(\ref{large-theta}). This limit also suggests a one-parameter fit of the full distribution, with the functional form 
\beq\label{fit}
\eta_{\textnormal{fit}}(z) =\frac{4c^{3/2}}{3 \sqrt{\pi}}\  \rme^{-c z^{2/3}}\,, 
\eeq
where the normalisation is such that Eq.~(\ref{eq:eta_norm}) is satisfied. For the fit-parameter we find $c\approx1.68$. For comparison we have plotted the fit $\eta_{\textnormal{fit}}(z)$ and also an exponential distribution $\eta_\text{exp}(z)=\rme^{-z}$ that has the first two moments identical to $\eta(z)$ in Fig.~\ref{fig2}.
The distributions have similar shapes, which comfort the choice of the exponential distribution in previous phenomenological studies of the missing energy in dijet events in Pb-Pb collisions \cite{Mehtar-Tani:2014yea,Blaizot:2014ula}.\\

Consider finally the hard part of the distribution. The single scattering limit, achieved when $\theta\gg \theta_\ast(x)$ can be recovered by Fourier transforming the first term in the series, $D_1(x,\u,t)$. One can also obtain it right away from Eq.~(\ref{evol-eq-ang2a}), which yields
\beq\label{D-it}
D_1(x,\btheta,L)= \int_0^L  \rmd t\int \frac{\rmd y}{y} D\left(\frac{x}{y},\frac{L-t}{\sqrt{y}}\right)\frac{x^2}{y^2}{\cal C}(\btheta,y) D(y,t)\,.
\eeq
Calculating the integral (\ref{D-it}) in the same way as we did for similar ones earlier, one easily obtains
 \beq
 D_1(x,\btheta,L)\approx D(x,L)\, {\cal C}(\btheta,x)\frac{t_\ast \sqrt{x}}{4}\approx D(x,L)\,\frac{\theta_\ast^2(x)}{\btheta^4}, \qquad \theta_\ast^2(x)= \frac{4 \pi^2\alpha^2_s N_c n t_\ast \sqrt{x}}{x^2 E^2}\,.
 \eeq
We can write 
\beq
\theta^2_\ast(x)\equiv\frac{1}{4\bar\alpha}\left[\frac{\hat q}{(xE)^3}\right]^{1/2}=\langle \theta^2\rangle\,.
\eeq
This is  the typical angle squared of gluons in the multiple branching regime. The prefactor $1/4$ is chosen to match the first moment, i.e., $\langle \theta^{2}\rangle= \theta^2_\ast(x)$. Note that this angle $\theta_\ast(x)$ is independent of the size of the medium.


\section{Summary}
\label{sec:summary}

In this work, we have constructed solutions to the equation that governs the evolution with the length of the medium of the inclusive gluon distribution along the medium-induced cascade that is initiated by a hard gluon of energy $E$. This equation resums multiple branchings and scatterings, which are responsible for the multiplication of partons along the cascade and for their momentum broadening.  We focused on the case where the energy of the leading particle is  large enough  so that it escapes the medium with a sizable fraction of its initial energy, that is,  the stopping time $t_\ast$ is larger than $L$, or equivalently  $E\gg \omega_s= \bar\alpha^2 \hat q L^2$.   This is the relevant condition to study the substructure of jets that are measured in Heavy-Ion collisions. In this case, one can identify three  distinct regimes in  the angular distribution, depending on the value of the energy $\omega$ of the energy of the observed gluon. 

The first regime, which reflects the property of the leading particle, is characterized by an energy $\omega\lesssim E $. The corresponding distribution shows a slight broadening in energy (towards lower energies) and angles (towards larger angles)  due to soft gluon emissions and  elastic kicks, respectively. These two effects are here independent of one another, which translates into the factorisation of the energy and the angular distributions.  The typical transverse momentum acquired by multiple scatterings in this regime is $
\langle k_\perp^2 \rangle = \hat q L$,  corresponding to the average angle squared $\langle \theta^2 \rangle = \hat q L/E^2$. 

The second regime corresponds to gluons in the cascade that have been primarily radiated by the leading partons and are to be found  in the  energy range $\omega_s\ll \omega\ll E$. The energy distribution is given by the BDMPSZ spectrum. These gluons can be emitted anywhere inside the medium and their angular broadening is determined  by  brownian motion which yields, on average
\beq
\langle \theta^2 \rangle = \theta^2_s(\omega,L)= \frac{\hat q L}{\omega^2}\,.
\eeq

In the regimes discussed so far,  the angular distribution is determined by processes that involve at most one gluon splitting. When $\omega\ll \omega_s$, multiple branching occur with probability one. This is the third regime. We already made the remark that the onset of the multitude branching regime at $\omega=\omega_s$ is not visible in  the energy distribution which exhibits a scaling behavior $\omega^{-1/2}$ for all $x\ll1$. But the character of the angular distribution changes at $\omega\lesssim \omega_s$. Indeed, in this regime, the characteristic squared angle is given by 
\beq
\theta^2_\ast(\omega) = \frac{\hat q t_\ast(\omega)}{\omega^2} = \frac{1}{\bar\alpha} \sqrt{ \frac{\hat q }{\omega^3}}\,.
\eeq
It is independent of the size of the medium: it corresponds to  the momentum broadening of the observed gluon, during the time $t_\ast(\omega)\ll L$ that this gluon spends in the medium from the moment it has been emitted. 

The characteristics times that have been recalled above, and which are related to average angle squared of the soft part of the angular distribution,  mark the frontier between this soft part of the distribution dominated by soft multiple scatterings,  and its hard tail dominated by hard scattering. This hard tail can be determined perturbatively at large angle by taking into account a single hard scattering.  The full angular distribution can be constructed  as a power series in the number of scatterings and computed numerically, as was shown in this paper. 

These results provide a first principle determination of the angular distribution of the gluons radiated by  a jet in a medium. They should be useful in future phenomenological studies of jet shapes.  \\

\noindent{\bf Acknowledgements}

We would like to thank F. Gelis for helpful discussions. LF acknowledges fruitful discussions with E. Iancu on related topics. 
This research is supported by the European Research Council under the Advanced Investigator Grant ERC-AD-267258.

\appendix

\section{Moments of the angular distribution in the diffusion approximation }
\label{sec:moments}

In this Appendix, we extend the calculation presented in Sect.~\ref{sec:angsquare} to the calculation of all the moments of the soft component of the angular distribution, that is, of the solution of Eq.~(\ref{evol-eq-diff0}).  
  We define normalized moments as follows
\beq\label{moments}
 \langle \theta^{2N}\rangle=\frac{M_N(x,L)}{M_0(x,L)}=\frac{\int_\btheta \btheta^{2N}\, D(x,\btheta,L)}{D(x,L)}\,,
\eeq
where $D(x,L)=\int_\btheta \, D(x,\btheta,L)$.
To obtain the equation satisfied by  $M_N(x,t)$, we multiply Eq.~(\ref{evol-eq-diff0}) by $\btheta^{2N}$, and integrate over $\btheta$. We obtain 
\beq\label{evol-eq-N}
\frac{\del}{\del t}M_N(x,t)=\frac{1}{t_\ast}\int \rmd z\, {\cal K}(z)\left[\sqrt{\frac{z}{x}} \,M_N\left(\frac{x}{z},t\right)-\frac{z}{\sqrt{x}} M_N\left(x,t\right)\right]+\frac{N^2\hat q}{(xE)^2} \,M_{N-1}(x,t)\,.
\eeq
The last term was obtained by integrating by parts over $\btheta$ in order to eliminate the Laplacian, and using the fact that all moments vanish at large $\btheta$. Eq.~(\ref{evol-eq-N}) is an inhomogeneous equation for the moment of order $N$, where the moment order $N-1$ plays the role of a source. This equation can be solved in the same way as Eq.~(\ref{evol-eq-1}), that is 
\beq\label{ang-sol}
M_N(x, L)=\frac{N^2 \hat q}{E^2}\int_0^L \rmd t\int_x^1 \frac{ \rmd y}{y^3}\, D\left(\frac{x}{y},\frac{L-t}{\sqrt{y}}\right)\, M_{N-1}(y,t)\,,
\eeq
with $D(x/y,(L-t)/\sqrt{y})$ given explicitly in Eq.~(\ref{Greenfunction}).

The ratios of moments, $r_N\equiv x^2 M_N/M_{N-1}$, obtained by solving numerically Eq.~(\ref{evol-eq-N}), are plotted in Fig.~\ref{fig2}. The limiting behaviors agree perfectly with the analytical analysis that we now turn to.
 \begin{figure}[!ht]
\begin{center}
\includegraphics[width=10cm]{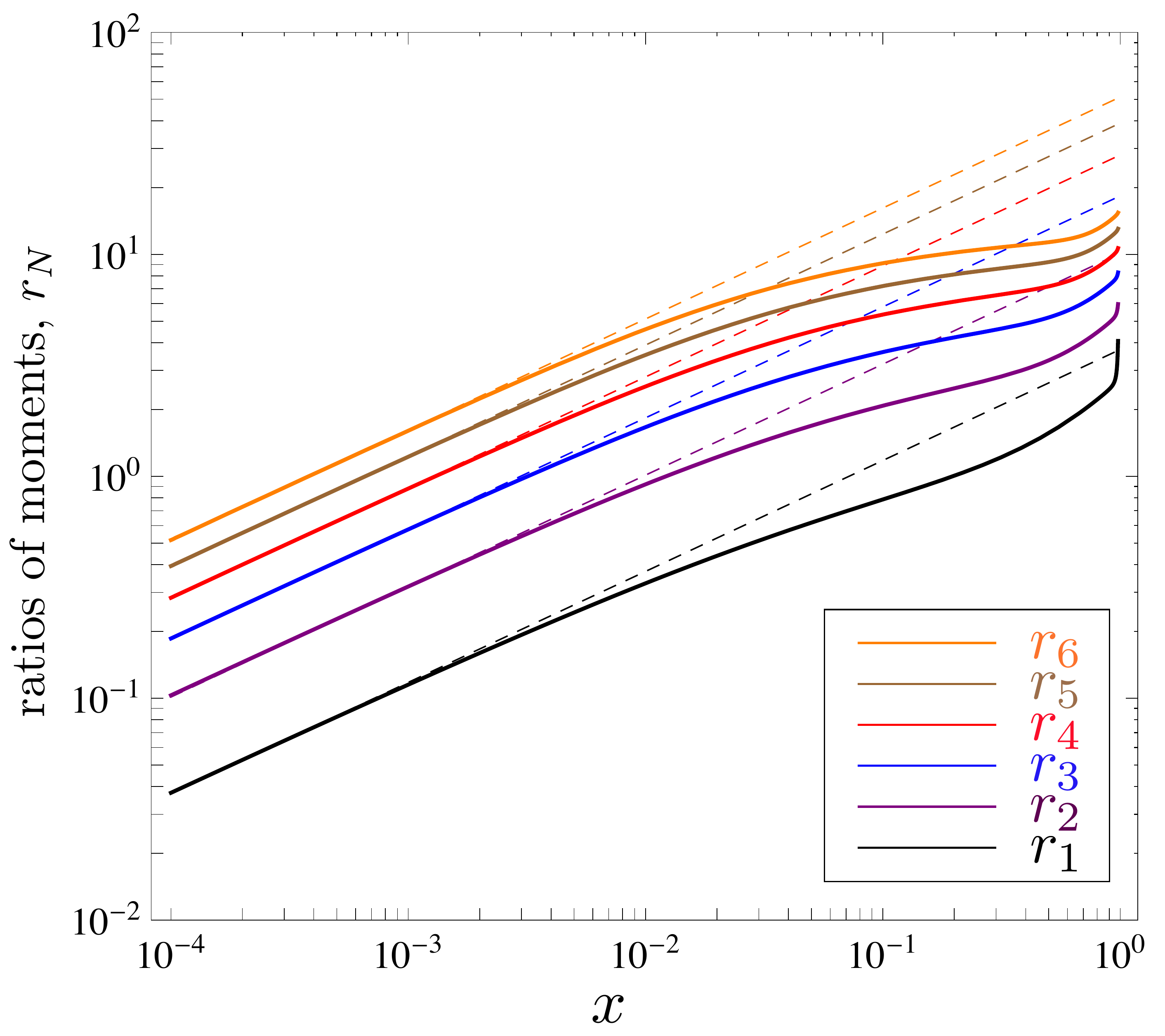}
\caption{(Color online.) The ratios of moments, $r_{N}\equiv x^2 M_N/M_{N-1}$ ($N=1\to 6$ from bottom to top), of the transverse momentum distribution, for the following set of parameters: $\hat q=1$ GeV$^2$/fm, $E=100$ GeV, $L=4$ fm, $\bar\alpha=0.3$. The small-$x$ limit extrapolations  are indicated by dashed lines. }

\label{fig3}
\end{center}
\end{figure}

\subsection{The region $x\lesssim 1$}
\label{sec:discussion}
When $x$ is very close to $1$, the angular distribution is not affected by gluon branching, and its soft component is given by the Gaussian (\ref{Gaussdistr2}), which is entirely characterized by the second moment $\langle\theta^2(x,L)\rangle=\theta^2_s(x,L)$.

\subsection{The region $x_s\ll x\ll 1$}
 
When  $x_s\ll x$ gluon branching can be taken into account perturbatively.   For $x$ not too small, the distribution remains determined by the first moment
 \beq
 M_1(x,L)=\frac{\hat q L}{2E^2}\left(1+\frac{1}{x^2}\right)D^{(1{\rm b})}(x,L)\,.
 \eeq
 When multiplied by $x^2E^2$ this moment represents the average momentum squared. The first contribution represents the average momentum square acquired by the gluon before its splitting, the second contribution  is that acquired after the splitting. The factor $1/2$ comes form the averaging over the time of the splitting in the interval $[0,L]$. Note that as $x$ becomes small, the first contribution becomes negligible: the entire contribution to the average momentum squared comes then from momentum broadening of the observed gluon since the time of its emission. We focus now on this contribution and write the first moment as
 \beq
 M_1(x,L)=\frac{\hat q L}{2x^2E^2}\,D^{(1{\rm b})}(x,L)\,.
 \eeq
More generally, by keeping the zero branching approximation for the Green's function in Eq.~(\ref{ang-sol}), we obtain a simple  recursion relation for the moments:
\beq
M_N(x,L)=\frac{N^2\hat q}{x^2 E^2}\int_0^L \rmd t\, M_{N-1}(x,t)\,.
\eeq
This can be solved in the form
\beq\label{MN2}
M_N(t) = a_N \,(\theta_s^{2}(x,t))^N \,D^{(1{\rm b})}(x,L)\,,
\eeq
where
\beq
 \theta^2_s(x,t)\simeq \frac{\hat q t}{2(xE)^2}=\frac{1}{2x^2}\frac{t}{L}\theta_s^2(L)\,.
\eeq
The recursion relation for the $a_N$ reads 
\beq
a_N = N^2\left(\int_0^1 \rmd u\, u^{N} \right) a_{N-1}=\frac{N^2}{N+1} \, a_{N-1}=\frac{N!}{N+1}\,, 
\eeq
so that the moments  take eventually the form
\beq
 M_N(L) = \frac{N!}{N+1} \,\frac{\theta_s^{2N}(L)}{x^{2N} }D^{(1{\rm b})}(x,L)\,.
\eeq
In particular, we have
\beq
r_N\equiv x^2\frac{M_N}{M_{N-1}}=\frac{N^2}{N+1} \theta_s^2=\frac{N^2}{N+1}\frac{\hat q L}{E^2}\,.
\eeq
This ratio is independent of $x$, a feature that can be recognized in Fig.~\ref{fig3} for the largest values of $N$ that are plotted.
The moments that we have obtained correspond to the moments of the angular distribution of the primarily emitted gluon (the BDMPSZ gluon), namely, 
\beq\label{modifiedexpo}
D(x,\btheta,L)= \int_0^L \rmd t \,\frac{4\pi}{ \theta^2_\text{s}(x,t) }\, \exp\left[ -\frac{\btheta^2}{ \theta^2_\text{s}(x,t)} \right]  \, D^{(1{\rm b})}(x,L)\,,
\eeq
with $D^{(1{\rm b})}(x,t)=(t/t_\ast)/\sqrt{x}$. It can indeed be verified that the moments of the distribution (\ref{modifiedexpo}) coincide with those given in Eq.~(\ref{MN2}).\\

\subsection{Small-$x$ limit}
\label{sec:late-times}

When $x\ll x_s$ multiple branchings are important and have to be fully taken into account. By using the same reasoning as that which leads to Eq.~(\ref{D-it1}), one easily obtain from  Eq.~(\ref{ang-sol}) 
\beq
M_N(x, L)=\frac{N^2\sqrt{ \hat qE}}{2\pi \bar\alpha E^2} \int_x^1 \frac{\rmd y}{y^3}\, \frac{\sqrt{y}}{\sqrt{x/y}\sqrt{1-x/y}}\, M_{N-1}(y,L)\,.
\eeq
This relation between moments confirms the fact that their time dependence is essentially unaltered as $N$ increases. In fact the entire $L$ dependence factorizes and drops in the ratio $M_N/M_0$ defining $\langle \theta^{2N}\rangle$ (see Eq.~(\ref{moments})). Recalling indeed that in the scaling region  $x,y\ll1$, we have  $D(y,L)/D(x,L)=\sqrt{x/y}$ we obtain the $L$-independent recursion formula for the moments
\beq\label{ang-sol-recurs}
\langle \theta^{2N}\rangle (x)=\frac{2 N^2}{\pi } \theta_\ast^2(x)\, \int_x^1 \rmd u \,\sqrt{\frac{u }{1-u } }\,\, \langle \theta^{2(N-1)}\rangle(x/u)\,,
\eeq
where $u=x/y$ and 
\beq
\theta^2_\ast(x)\equiv\frac{1}{4\bar\alpha}\left[\frac{\hat q}{(xE)^3}\right]^{1/2}=\langle \theta^2\rangle\,,
\eeq
is the typical angle squared of gluons in the multiple branching regime. The prefactor $1/4$ is chosen to match the first moment, i.e., $\langle \theta^{2}\rangle= \theta^2_\ast(x)$. Note that this angle $\theta_\ast(x)$ is independent of the size of the medium.

In order to solve the recursion relation (\ref{ang-sol-recurs}) we set
\beq
\langle \theta^{2N}\rangle =  a_N \,\theta^{2N}_\ast(x)\,.
\eeq
We then obtain  the following equation for the coefficients $a_N$ : 
\beq\label{a-N-coef}
a_N =\frac{2N^2}{\pi } \, \int_0^1 \rmd u \frac{u^{\frac{3N-2}{2}} }{\sqrt{1-u} }\, a_{N-1}= \frac{2 N^2}{\sqrt{\pi }}   \frac{\Gamma\left(\frac{3N}{2}\right)} {\Gamma\left(\frac{1+3N}{2}\right)}\, a_{N-1}\,.
\eeq
The first four coefficients of the moments are, 
\beq
a_0=1,\qquad  a_1=1,\qquad  a_2=\frac{128}{15\pi},\qquad  a_3=\frac{42}{\pi}\,.
\eeq
These differ  somewhat from the corresponding values $a_N=N!$ for the exponential distribution.
In fact, at large $N$,  Stirling formula ($\Gamma(x)\sim x^{-x}$) allows us to write 
\beq\label{as-mom1}
a_N\sim  N^{\frac{3N}{2}}\sim \Gamma\left(\frac{3N}{2}\right)\,.
\eeq
These values are those corresponding to a distribution that falls like $e^{-c\, \theta^{\beta}}$,  
\beq\label{as-mom2}
a_N = \int \rmd \theta^2\, \theta^{2N}\, \rme^{-c \,\theta^\beta } \sim \Gamma\left(\frac{2 N}{\beta}\right)\,,
\eeq
with  $\beta=4/3$. This suggests that the soft part of the angular distribution  behaves at large  angles  as 
\beq\label{large-theta}
D(x, \theta,L) \sim \rme^{-c \,\theta^{4/3}}\,.
\eeq
Of course, this asymptotic behavior is only that of the soft part of the distribution, for which moments exist. It is visible in Fig.~\ref{fig2}, but will be hidden by that actual tail corresponding to single scatterings.

\providecommand{\href}[2]{#2}\begingroup\raggedright

\endgroup
\end{document}